\documentclass[12pt,preprint]{aastex}

\begin{document}

\title{\large A Standard Kinetic Energy Reservoir in Gamma-Ray Burst
Afterglows}

\author{E. Berger\altaffilmark{1} and S. R. Kulkarni\altaffilmark{1}}
\affil{Division of Physics, Mathematics, and Astronomy,
California Institute of Technology 105-24, Pasadena, CA 91125}
\email{ejb@astro.caltech.edu; srk@astro.caltech.edu}
\author{D.~A. Frail\altaffilmark{2}}
\affil{National Radio Astronomy Observatory, Socorro, NM 87801}
\email{dfrail@nrao.edu}

\begin{abstract} 

We present a comprehensive sample of X-ray observations of 41
$\gamma$-ray burst (GRB) afterglows, as well as jet opening angles,
$\theta_j$ for a subset with measured jet breaks.  We show that there
is a significant dispersion in the X-ray fluxes, and hence isotropic
X-ray luminosities ($L_{X,{\rm iso}}$), normalized to $t=10$ hr.
However, there is a strong correlation between $L_{X,{\rm iso}}$ and
the beaming fractions, $f_b\equiv[1-{\rm cos}(\theta_j)]$.  As a
result, the true X-ray luminosity of GRB afterglows, $L_X=f_bL_{X,{\rm
iso}}$, is approximately constant, with a dispersion of only a factor
of two.  Since $\epsilon_eE_b\propto L_X$, the strong clustering of
$L_X$ directly implies that the adiabatic blastwave kinetic energy in
the afterglow phase, $E_b$, is tightly clustered.  The narrow
distribution of $L_X$ also suggests that $p\approx 2$, that inverse
Compton emission does not in general dominate the observed X-ray
luminosity, and that radiative losses at $t<10$ hr are relatively
small.  Thus, despite the large diversity in the observed properties
of GRBs and their afterglows the energy imparted by the GRB central
engine to the relativistic ejecta is approximately constant.

\end{abstract}


\section{Introduction}
\label{sec:introduction}

Gamma-ray bursts (GRBs) exhibit a remarkable diversity: Fluences range
from $10^{-7}$ to $10^{-3}\,$erg cm$^{-2}$, peak energies range from
50\,keV to an MeV, and possibly from the X-ray to the GeV band
\citep{fm95}, and durations extend from about 2 to $10^3$ s (for the
long-duration GRBs).  This diversity presumably reflects a dispersion
in the progenitors and central engine properties.  Perhaps the most
impressive feature of GRBs are their brilliant luminosities and
isotropic energy releases approaching the rest mass of a neutron star,
$E_{\gamma,\rm iso}\sim 10^{54}\,$ erg \citep{kdo+99,ahp+00}.  

The quantity of energy imparted to the relativistic ejecta, $E_{\rm
rel}$, and the quality parameterized by the bulk Lorentz factor,
$\Gamma$, are the two fundamental properties of GRB explosions.  In
particular, extremely high energies push the boundaries of current
progenitor and engine models, while low energies could point to a
population of sources that is intermediate between GRBs and
core-collapse supernovae.   

The true energy release depends sensitively on the geometry of the
ejecta.  If GRB explosions are conical (as opposed to spherical) then
the true energy release is significantly below that inferred by
assuming isotropy.  Starting with GRB\,970508 \citep{wkf98,rho99}
there has been growing observational evidence for collimated outflows,
coming mainly from achromatic breaks in the afterglow lightcurves.

In the conventional interpretation, the epoch at which the afterglow
lightcurves steepen (``break'') corresponds to the time at which
$\Gamma$ decreases below
$\theta_j^{-1}$, the inverse opening angle of the collimated outflow
or ``jet'' \citep{rho99}.  The break happens for two reasons: an edge
effect, and lateral spreading of the jet which results in a
significant increase of the swept up mass.  Many afterglows have $t_j 
\sim 1-{\rm few}$ days, which are best measured from optical/near-IR
lightcurves (e.g.~\citealt{hbf+99,kdo+99,sgk+99}), while wider opening
angles are easily measured from radio lightcurves
(e.g.~\citealt{wkf98,bdf+01}).   

Recently, \citet{fks+01} inferred $\theta_j$ for fifteen GRB
afterglows from measurements of $t_j$ and found the surprising result
that $E_{\gamma,\rm iso}$ is strongly correlated with the beaming
factor, $f_b^{-1}$; here, $f_b\equiv[1-{\rm cos}(\theta_j)]$ is the
beaming fraction and $E_{\gamma,\rm iso}$ is the $\gamma$-ray energy
release inferred by assuming isotropy.  In effect, the true
$\gamma$-ray energy release, $E_\gamma=f_bE_{\gamma,{\rm iso}}$
is approximately the same for all the GRBs in their sample, with a
value of about $5\times 10^{50}$ erg (assuming a constant circumburst 
density, $n_0=0.1$ cm$^{-3}$).  In the same vein, broad-band modeling 
of several GRB afterglows indicates that the typical blastwave kinetic
energy in the adiabatic afterglow phase is $E_b\sim 5\times
10^{50}$ erg, with a spread of about 1.5 orders of magnitude
\citep{pk02}. However, the general lack of high quality afterglow
data severely limits the application of the broad-band modeling
method. 

Separately, \citet{kum00} and \citet{fw01} noted that the afterglow
flux at frequencies above the synchrotron cooling frequency, $\nu_c$,
is proportional to $\epsilon_e dE_b/d\Omega$, where $\epsilon_e$ is
the fraction of the shock energy carried by electrons and $dE_b/
d\Omega$ is the energy of the blastwave per unit solid angle.
The principal attraction is that the flux above $\nu_c$ does not
depend on the circumburst density, and depends only weakly on the
fraction of shock energy in magnetic fields, $\epsilon_B$.  For
reasonable conditions (which have been verified by broad-band
afterglow modeling, e.g.~\citealt{pk02}), the X-ray band ($2-10$ keV)
lies above $\nu_c$ starting a few hours after the burst.  Thus, this
technique offers a significant observational advantage, namely the
X-ray luminosity can be used as a surrogate for the
isotropic-equivalent afterglow kinetic energy.   

\citet{pkp+01} find that the X-ray flux, estimated at a common epoch
($t=11\,$hr), exhibits a narrow distribution of $\log(F_X)$, $\sigma_l
(F_X) = 0.43^{+0.12}_{-0.11}$; here $\sigma_l^2(x)$ is the variance of
$\log(x)$.  Taken at face value, the narrow distribution of $F_X$
implies a narrow distribution of $\epsilon_edE_b/d\Omega$.  This
result, if true, is quite surprising since if the result of
\citet{fks+01} is accepted then $dE_b/d\Omega$ should show a wide
dispersion comparable to that of $f_b^{-1}$.

\cite{pkp+01} reconcile the two results by the following argument.
The relation between $dE_b/d\Omega$ and $E_b$ can be restated as
$\log(dE_b/d\Omega) = \log(E_b) + \log(f_b^{-1})$.  Thus, $\sigma_l^2
(dE_b/d\Omega) = \sigma_l^2(E_b) +  \sigma_l^2(f_b^{-1})$.  Since
$dE_b/d\Omega\propto L_{X,\rm iso}$ (for a constant $\epsilon_e$) they
express, $\sigma_l^2(E_b)=\sigma_l^2(L_{X,\rm iso})-\sigma_l^2(f_b)$.
Given the diversity in $\theta_j$ \citep{fks+01} and the apparent
narrowness in $F_X$ (above), it would then follow that $E_b$ should be
very tightly distributed. 

However, the approach of \citet{pkp+01} makes a key assumption, namely
that $E_b$ and $f_b^{-1}$ are uncorrelated.  This is certainly true
when $E_b$ is constant, but the assumption then pre-supposes the
answer!  In reality, a correlation between $E_b$ and $f_b$ can either
increase or decrease $\sigma_l^2(E_b)$, and this must be addressed
directly.  Finally, as appears to be the case (see \S\ref{sec:data}),
$\sigma_l^2(f_b^{-1}$) is dominated by bursts with the smallest
opening angles, which results in a distinctly different value than the
one used by \citet{pkp+01} based only on the observed $\theta_j$
values. 

In this {\it Letter}, we avoid all these concerns by taking a direct
approach: we measure the variance in $E_b\propto f_b L_{X,\rm iso}$
rather than bounding it through a statistical relation.  We show, with
a larger sample, that $L_{X,\rm iso}$ is not as narrowly distributed
as claimed by \citet{pkp+01}, and in fact shows a spread similar to
that of $E_{\gamma,\rm iso}$.  On the other hand, we find that
$L_{X,{\rm iso}}$ is strongly correlated with $f_b^{-1}$.  It is this
correlation, and not the claimed clustering of $L_{X,\rm iso}$, that
results in, and provides a physical basis for the strong clustering of
$L_X$ and hence the blastwave kinetic energy, $E_b$.

\section{X-ray Data}
\label{sec:data}

In Table~\ref{tab:xray} we provide a comprehensive list of X-ray
observations for 41 GRB afterglows, as well as temporal decay indices,
$\alpha_X$ ($F_\nu\propto t^{\alpha_X}$), when available.  In
addition, for a subset of the afterglows for which jet breaks have
been measured from the radio, optical, and/or X-ray emission, we also
include the inferred $\theta_j$ \citep{fks+01,pk02}.  We calculate
$\theta_j$ from $t_j$ using the circumburst densities inferred from
broad-band modeling, when available, or a fiducial value of $10$
cm$^{-3}$, as indicated by the best-studied afterglows
(e.g.~\citealt{yfh+02}).  This normalization for $n_0$ is different
from \citet{fks+01} who used $n_0=0.1$ cm$^{-3}$.

For all but one burst we interpolate the measured $F_X$ to a fiducial
epoch of 10\,hr (hereafter, $F_{X,10}$), using the measured $\alpha_X$
when available, and the median of the distribution, $\langle\alpha_X
\rangle=-1.33\pm 0.38$ when a measurement is not available.  The
single exception is GRB\,020405 for which the first measurement was
obtained $t\approx 41$ hr, while the inferred jet break time is about
$23$ hr (Berger et al.~in prep).  In this case, we extrapolate to $t=10$ hr
using $\alpha_X=-1.69$ for $t>23$ hr and $\alpha_X=-0.78$ for $t<23$
hr.  We list the values of $F_{X,10}$ in Table~\ref{tab:xray2}. 

\begin{figure} 
\epsscale{0.8}
\plotone{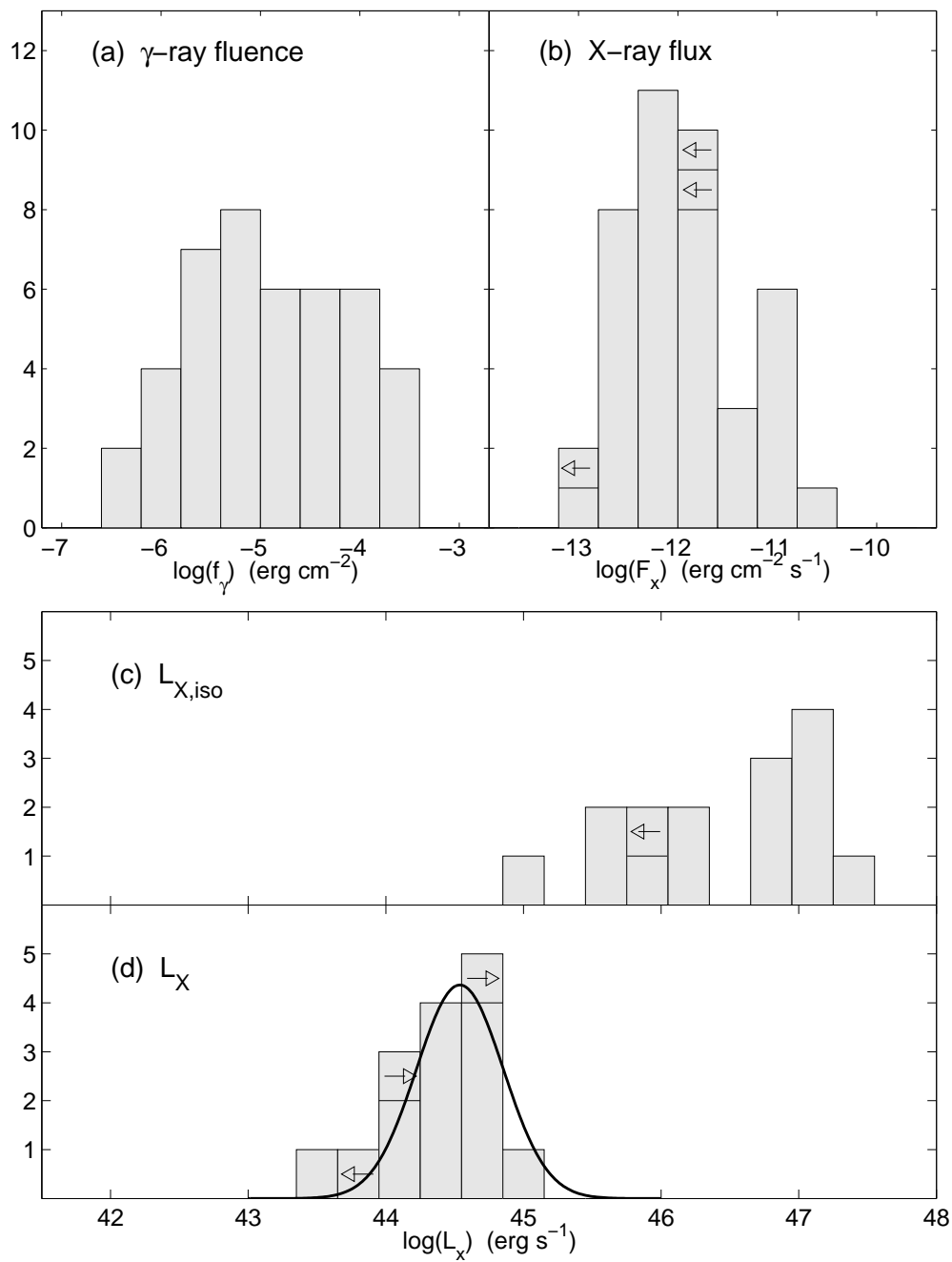}
\caption{Panel (a) shows the distribution of $\gamma$-ray fluences.
Panel (b) shows the distribution of X-ray fluxes scaled to $t=10$ hr
after the burst.  In panel (c) we plot the isotropic-equivalent X-ray
luminosity, $L_{X,{\rm iso}}$, for the subset of X-ray afterglows with
known $\theta_j$ and redshift, while in panel (d) we show the true
X-ray luminosity, $L_X=f_b^{-1}L_{X,{\rm iso}}$.
\label{fig:fluxes}}
\end{figure}

In Figure~\ref{fig:fluxes} we plot the resulting distribution of 
$F_{X,10}$.  For comparison we also show the distribution of
$\gamma$-ray fluences from the sample presented by \citet{bfs01}
and updated from the literature.  Clearly, while the distribution of
X-ray fluxes is narrower than that of the $\gamma$-ray
fluences, $\sigma_l(f_\gamma)= 0.79^{+0.10}_{-0.08}$, it still
spans $\sim 2.5$ orders of magnitude, i.e.~$\sigma_l(F_{X,10})=
0.57^{+0.07}_{-0.06}$.  The value of $\sigma_l(F_{X,10})$, and all
variances quoted below, are calculated by summing the Gaussian
distribution for each measurement, and then fitting the combined
distribution with a Gaussian profile. 

We translate the observed X-ray fluxes to isotropic luminosities
using: 
\begin{equation}
L_{X,{\rm iso}}(t=10\,{\rm hr})=4\pi d_L^2 F_{X,10}
(1+z)^{\alpha_X-\beta_X-1}.
\end{equation}
We use $\beta_X\approx -1.05$, the weighted mean value for X-ray
afterglows \citep{dpp+02}, and the median redshift, $\langle
z\rangle=1.1$, for bursts that do not have a measured redshift.  The
resulting distribution of $L_{X,{\rm iso}}$, $\sigma_l(L_{X,{\rm
iso}})=0.68^{+0.17}_{-0.09}$, is wider than that of $F_X$ due
to the dispersion in redshifts.  We note that this is significantly
wider than the value quoted by \citet{pkp+01} of $\sigma_l(L_{X,{\rm
iso}})\approx 0.43$ based on a smaller sample.  Using the same method
we find $\sigma_l(E_{\gamma,{\rm iso}})=0.92^{+0.12}_{-0.08}$.  

In the absence of a strong correlation between $f_b$ and $L_{X,{\rm
iso}}$, the above results indicate that the distribution of the true
X-ray luminosities, $L_X\equiv f_b^{-1}L_{X,{\rm iso}}$, should have a
wider dispersion than either $L_{X,\rm iso}$ or $f_b$, for which we
find $\sigma_l(f_b)=0.52^{+0.13}_{-0.12}$ \citep{fks+01}.  Instead,
when we apply the individual beaming corrections for those bursts that
have a measured $\theta_j$ and
redshift\footnotemark\footnotetext{These do not include GRB\,990705
which is poorly characterized; see \S\ref{sec:energy}.} (see
Table~\ref{tab:xray2}), we find a significantly narrower distribution,
$\sigma_l(L_X)=0.32^{+0.10}_{-0.06}$.

\section{Beaming Corrections and Kinetic Energies}
\label{sec:energy}

\begin{figure}
\epsscale{0.8} 
\plotone{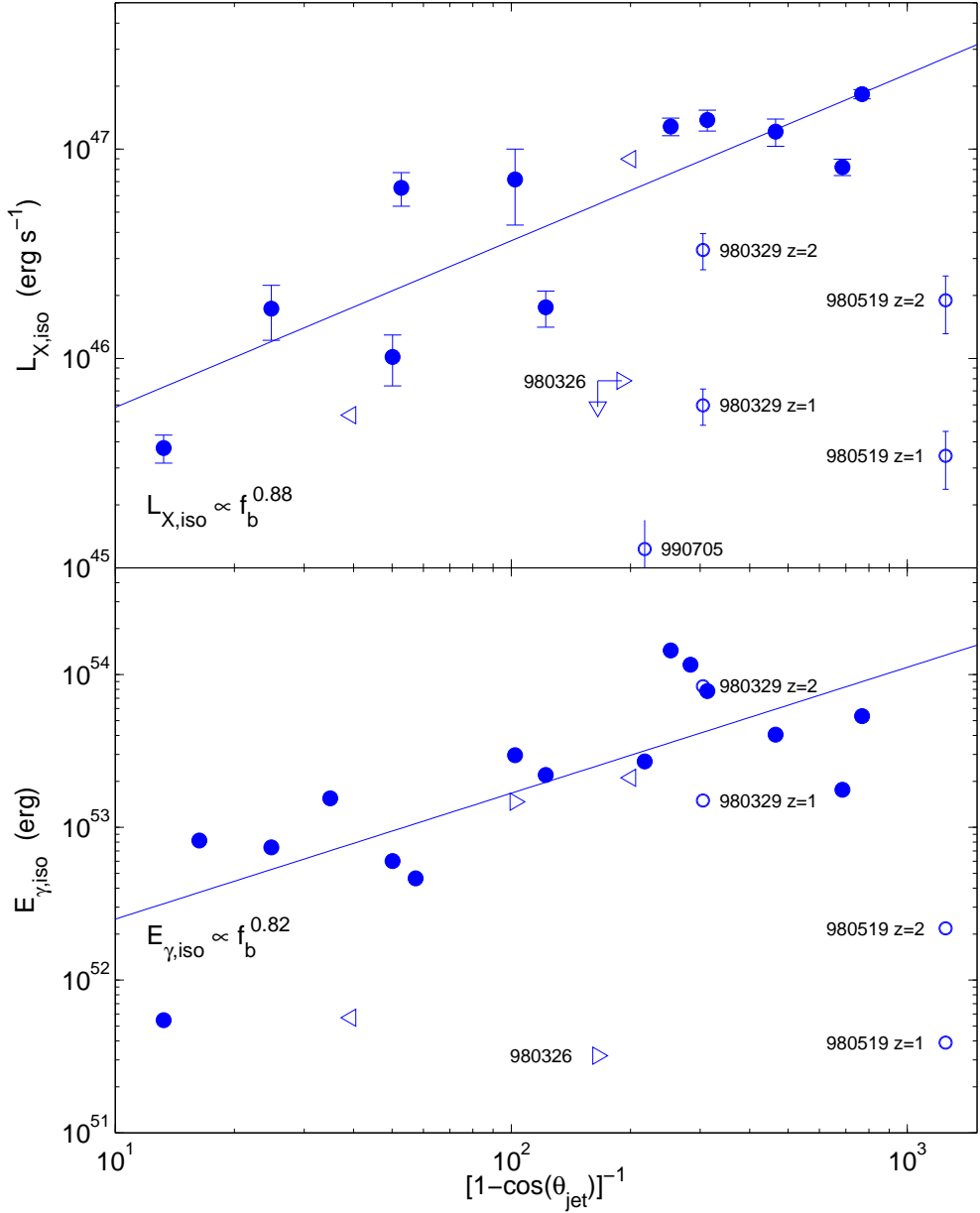}
\caption{Isotropic-equivalent X-ray luminosity (top) and
isotropic-equivalent $\gamma$-ray energy (bottom) as a function of the
beaming factor, $[1-{\rm cos}(\theta_j)]^{-1}$.  There is a strong
positive correlation between $L_{X,{\rm iso}}$ and $f_b^{-1}$, as
well as between $E_{\gamma,{\rm iso}}$ and $f_b^{-1}$ resulting in an
approximately constant true X-ray luminosity and $\gamma$-ray energy
release.  In fact, while the distributions of all three parameters
span about three orders of magnitude, the distributions of the
beaming-corrected parameters span about one order of magnitude.
\label{fig:theta}}
\end{figure}

The reduced variance of $L_X$ compared to that of $L_{X,\rm iso}$
requires a strong correlation between $L_{X,{\rm iso}}$ and
$f_b^{-1}$, such that bursts with a brighter isotropic X-ray
luminosity are also more strongly collimated.  Indeed, as can be seen
from Figure~\ref{fig:theta} the data exhibit such a correlation.
Ignoring the two bursts which are obvious outliers (980326 and
990705), as well as GRBs 980329 and 980519, which do not have a
measured redshift, we find $L_{X,{\rm iso}}\propto f_b^{0.88}$.  The
linear correlation coefficient between $L_{X,{\rm iso}}$ and $f_b$
indicates a probability that the two quantities are not correlated of
only $4.6\times 10^{-4}$.  For $E_{\gamma,{\rm iso}}$ and $f_b$ we
find a similar probability of $4.2\times 10^{-4}$ that the two
quantities are not correlated. 

Thus, as with the $\gamma$-ray emission, the afterglow emission also
exhibits strong luminosity diversity due to strong variations in
$f_b$.  Therefore, the mystery of GRBs is no longer the energy release
but understanding what aspect of the central engine drives the wide
diversity of $f_b$. 

We note that there are four possible outliers in the correlation
between 
$L_{X,{\rm iso}}$ and $f_b^{-1}$.  The afterglows of GRBs 980326 and
980519 exhibit rapid fading \citep{ggv+98,vhc+00}, which has been
interpreted as the signature of an early jet break.  However, it is
possible that the rapid fading is instead due to a $\rho\propto
r^{-2}$ density profile, and in fact for GRB\,980519 such a model
indicates $\theta_j\approx 0.12$, $3$ times wider than in the constant
density model.  This is sufficient to bring GRB\,980519 into agreement
with the observed correlation.  The redshift of GRB\,980329 is not
known, but with $z=2$ it easily agrees with the correlation.  Finally,
the X-ray flux and jet opening angle for GRB\,990705 are poorly
characterized due to contamination from a nearby source \citep{dpp+02}
and a poor optical lightcurve \citep{mpp+00}.

\section{Discussion and Conclusions}
\label{sec:dis}

We have presented a comprehensive compilation of early X-ray
observations of 41 GRBs, from which we infer $F_{X,10}$, the flux
in the 2--10\,keV band at 10\,hr.  As first pointed by \citet{kum00} 
and \citet{fw01}, the afterglow luminosity above the cooling frequency
is $L_{X,iso}\propto\epsilon_e E_{b,\rm iso}$ where $E_{b,\rm iso}$
is the isotropic-equivalent explosion kinetic energy.  More
importantly, the flux is independent of the ambient density and weakly
dependent on $\epsilon_B$. For all well modeled afterglows, the
cooling frequency at 10\,hr is below the X-ray band.  Thus, $F_{X,10}$
can be utilized to yield information about the kinetic energy of GRBs.

Earlier work \citep{pkp+01} focussed on statistical studies of
$F_{X,10}$ and found the very surprising result that it is narrowly
clustered.  By assuming that the true kinetic energy, $E_b=E_{b,\rm
iso}f_b\propto L_X = L_{X,\rm iso}f_b$, and $f_b$ (the beaming factor)
are uncorrelated, the authors deduced that $L_X$ and thus $E_b$ are
even more strongly clustered.  However, this approach is weakened by
assuming (in effect) the answer.  Furthermore, the approach of
\citet{pkp+01} which relies on subtracting variances is very sensitive
to measurement errors.  To illustrate this point, we note
$\sigma_l^2(L_{X,\rm iso})=0.68^{+0.17}_{-0.09}$ for the entire sample
presented here, whereas $\sigma_l^2(f_b)=0.52^{+0.13}_{-0.12}$.  Thus,
$\sigma^2_l(L_X)=0.16^{+0.30}_{-0.21}$ may be negative using the
statistical approach.

In contrast to the statistical approach, we take the direct approach
and estimate the true kinetic energy, $E_b \propto L_{X,\rm iso}f_b$,
by using the measured $L_{X,\rm iso}$ and inferred $f_b$.  The
advantage of our approach is that we do not make assumptions of
correlations (or lack thereof) and more importantly we do not subtract
variances.  We directly compute the variance of the desired physical
quantity, namely $L_X$, and find that it is strongly clustered.  

Even more importantly, with our direct approach we have uncovered the
physical reason for the wide dispersion in $L_{X,{\rm iso}}$ and the
clustering of $L_X$, namely the dispersion in jet opening angles.   

$L_X$ is related to the physical quantities as follows 
\citep{fw01}: 
\begin{equation}
\epsilon_eE_b \propto A L_X Y^\epsilon, 
\label{eqn:Lx-Eb}
\end{equation} 
where
\begin{equation}
Y \equiv B \epsilon_e^{-3} \epsilon_B^{-1} L_{X,\rm iso}^{-1}.
\label{eqn:Y}
\end{equation}
Here $\epsilon\equiv(p-2)/(p-1)$, as well as $A$ and $B$ depend to
some extent on the details of the electron distribution (power law
versus relativistic Maxwellian; the value of power law index, $p$). 

There is no reason to expect that $L_X$ should be clustered.  However,
one can argue that the microphysics should be the same for each GRB
afterglow, in particular $\epsilon_e$ and $p$.  The best studied
afterglows appear to favor $p=2.2$ (e.g.~\citealt{fwk00,gwb+98}), 
a value also favored by our current theoretical knowledge of shock
acceleration (see~\citealt{ob02} and references therein).  In
addition, as already indicated by the $\gamma$-ray observations, there
is evidence supporting strong clustering of explosion energies in GRBs
\citep{fks+01}. 

Given these reasonable assumptions, a strong clustering of $L_X$ makes
sense if the physical quantities that are responsible for $L_X$ are
clustered.  As can be seen from Equation~\ref{eqn:Lx-Eb}, this would
require that $L_X$ be linearly related to $E_b$.  Such a relation is
possible if three conditions are met.  

First, the afterglow X-ray emission on timescales of 10 hr  must be
primarily dominated by synchrotron emission (which is the basis of
Equation~\ref{eqn:Lx-Eb}).  Contribution from inverse Compton (IC)
emission, which depends strongly on $n_0$ and $\epsilon_B$
\citep{se01}, is apparently not significant.  A possible exception is
GRB\,000926 \citep{hys+01}, but even there the IC contribution is
similar to that from synchrotron emission.  

Second, the energy radiated by the afterglow from the time of the
explosion to $t=10$ hr cannot be significant.  This constrains the
radiative losses at early time to at most a factor of few.

Third, $p$ must be relatively constant (as one may expect in any case 
from insisting that the microphysics should not be different for
different bursts).  For example, changing $p$ from a value of $1.5$ to
$3$ results in $Y^\epsilon$ ranging from 0.003 to 117, a factor of
39,000!  Even small changes in $p$, e.g.~from $p=1.75$ to $p=2.25$,
result in a factor of 8 change in $Y^\epsilon$.  In contrast, some
afterglow models yield values of $p$ significantly below $2$
(e.g.~\citealt{pk02}), while others have $p$ approaching $3$
\citep{cl00}.  Our results, on the other hand, indicate that one
should set $p\approx 2$ and attribute apparent deviant values of $p$
to external environment or energy injection from the central source. 

We end with an interesting conclusion from the results presented
here.  Since both the prompt and afterglow emission exhibit a strong
correlation with $f_b$, which is determined from late-time
observations (hours to weeks after the burst), the resulting constancy
of both $E_\gamma$ and $E_b$, indicates that GRB jets must be
relatively homogeneous and maintain a simple conical geometry all the
way from internal shocks ($\sim 10^{13}-10^{14}$ cm) to the epoch of
jet break ($\sim 10^{17}$ cm).  This rules out the idea that brighter
bursts are due to bright spots along specific lines of sight
\citep{kp00}, or that GRB jets have a strong energy and/or Lorentz
factor gradient across their surface \citep{rlr02}.  It is indeed
remarkable that the simplest description of jets is fully consistent
with the observations.

\acknowledgements

SRK thanks S.~Phinney for valuable discussions.  We acknowledge
support from SNF and NASA grants.


\clearpage

\begin{deluxetable}{ccccccc}
\tablecolumns{7}
\tablewidth{0pc}
\tablecaption{X-ray Afterglow Data\label{tab:xray}} 
\tablehead {
\colhead {GRB}        	&
\colhead {$z$} 		&
\colhead {Epoch}	&
\colhead {Flux}		&
\colhead {$\alpha_X$}	&
\colhead {$\theta_{\rm jet}$}	&
\colhead {Ref.}		\\
\colhead {}		&
\colhead {}		&
\colhead {(hrs)}	&
\colhead {($10^{-13}$ erg cm$^{-2}$ s$^{-1}$)}	&
\colhead {}		&
\colhead {}		&
\colhead {}		
}
\startdata
970111 & \nodata  & 24.0 & $1.05\pm 0.46$ & $-0.4\pm 3.2^a$ & \nodata & 1,2 \\ 
       &       & 30.7 & $0.95\pm 0.34$ &                 & \nodata & 2 \\
970228 & 0.695 & 8.5  & $33.8\pm 3.3$  & $-1.27\pm 0.14$ & \nodata & 2,3 \\ 
       &       & 12.7 & $28\pm 4$      &                 &         & 2 \\
       &       & 92.4 & $1.5\pm 0.4$   &                 &         & 2 \\ 
970402 & \nodata & 9.9  & $2.9\pm 0.4$   & $-1.35\pm 0.55$ & \nodata & 2 \\ 
       &       & 16.8 & $1.5\pm 0.4$   &                 &         & 2 \\ 
970508 & 0.835 & 13.1 & $7.13$         & $-1.1\pm 0.1$   & 0.391   & 4,5 \\
       &       & 72.3 & $4.3\pm 0.5$   &                 &         & 2 \\ 
       &       & 104  & $2.3\pm 0.7$   &                 &         & 2 \\ 
970815 & \nodata & 89.6 & $<1$           & \nodata         & \nodata & 6 \\
970828 & 0.958 & 4.0  & $118$          & $-1.42$         & 0.128   & 5,7 \\
       &       & 42.6 & $4.1$          &                 &         & 7 \\
971214 & 3.418 & 8.1  & $9.0\pm 0.9$   & $-1\pm 0.2$     & $>0.100$ & 2,5 \\ 
       &       & 28.9 & $2.1\pm 0.4$   &                 &         & 2 \\ 
971227 & \nodata & 16.5 & $2.5\pm 0.7$   & $-1.12\pm 0.06$ & \nodata & 8 \\
980326 & $\sim 1^b$ & 8.5  & $<16$     & \nodata         & $<0.110$ & 9 \\
980329 & \nodata & 8.4  & $14\pm 2.1$  & $-1.55\pm 0.3$  & 0.081   & 10,11 \\
       &       & 11.8 & $6.2\pm 1.2$   &                 &         & 10 \\
       &       & 16.4 & $3.4\pm 1.0$   &                 &         & 10 \\
       &       & 23.7 & $2.7\pm 0.7$   &                 &         & 10 \\
       &       & 43.6 & $1.1\pm 0.4$   &                 &         & 10 \\
980515 & \nodata & 11 & $2.0^{+0.5}_{-0.9}$ & \nodata    & \nodata & 12 \\
980519 & $<2^c$ & 10.9 & $5.3\pm 1.0$   & $-1.7\pm 0.7$   & 0.040   & 13,14 \\
       &       & 15.3 & $2.0\pm 0.4$   &                 &         & 13 \\
       &       & 21.5 & $1.6\pm 0.5$   &                 &         & 13 \\
       &       & 27.2 & $0.8\pm 0.4$   &                 &         & 13 \\
980613 & 1.096 & 9.9  & $7.1\pm 1.9$   & $-0.92\pm 0.62$ & $>0.226$ & 2 \\ 
       &       & 23.4 & $4.0\pm 0.8$   &                 &         & 2 \\ 
980703 & 0.966 & 34.0 & $4.0\pm 1$     & $-1.24\pm 0.18$ & 0.200   & 2,15 \\
981226 & \nodata & 14.0 & $4.0$	       & $-1.3\pm 0.4$   & \nodata & 16 \\
990123 & 1.600 & 6.4  & $124\pm 11$    & $-1.41\pm 0.05$ & 0.089   & 2,5 \\ 
       &       & 23.4 & $19.1\pm 2.2$  &                 &         & 2 \\ 
990217 & \nodata & 11   & $<1.1$         & \nodata         & \nodata & 12 \\
990510 & 1.619 & 8.7  & $47.8\pm 3.1$  & $-1.41\pm 0.18$ & 0.054   & 5,14,17 \\
       &       & 10.1 & $40.5\pm 2.6$  &                 &         & 17 \\
       &       & 11.7 & $32.8\pm 3.7$  &                 &         & 17 \\
       &       & 13.4 & $22.8\pm 2.8$  &                 &         & 17 \\
       &       & 15.3 & $24.1\pm 2.7$  &                 &         & 17 \\
       &       & 17.1 & $18.5\pm 3.1$  &                 &         & 17 \\
       &       & 19.1 & $20.9\pm 2.3$  &                 &         & 17 \\
       &       & 24.0 & $12.1\pm 1.4$  &                 &         & 17 \\
       &       & 26.3 & $9.9\pm 1.1$   &                 &         & 17 \\
       &       & 29.4 & $7.8\pm 1.1$   &                 &         & 17 \\
990627 & \nodata & 11.9 & $3.5$        & \nodata         & \nodata & 18 \\
990704 & \nodata & 10.1 & $10.1\pm 2.9$  & $-1.3\pm 0.3$   & \nodata & 19 \\
       &       & 13.4 & $8.9\pm 2.2$   &                 &           & 19 \\
       &       & 23.3 & $3.1\pm 2.0$   &                 &           & 19 \\
       &       & 26.8 & $2.9\pm 1.6$   &                 &           & 19 \\
990705 & 0.840 & 14.5 & $1.9\pm 0.6$   & \nodata         & 0.096   & 5,20 \\
990806 & \nodata & 13.6 & $5.5\pm 1.5$   & $-1.4\pm 0.7$   & \nodata & 21 \\
       &       & 34.3 & $1.5\pm 0.6$   &	         &           & 21 \\
990907 & \nodata & 11 & $10.2\pm 5.6$  & \nodata         & \nodata & 12 \\
991014 & \nodata & 11 & $4.0^{+1.4}_{-1.2}$ & \nodata    & \nodata & 12 \\
991216 & 1.020 & 4.0  & $1240\pm 40$   & $-1.61\pm 0.07$ & 0.051   & 5,14,22 \\
       &       & 10.9 & $250\pm 10$    &		 &         & 22 \\
000115 & \nodata & 2.9  & $270$          & $<-1$           & \nodata & 23 \\
000210 & 0.846 & 11   & $4.0\pm 1.0$   & $-1.38\pm 0.03$ & \nodata & 24 \\	
000214 & \nodata & 14.9 & $5$            & $-1.8$          & \nodata & 25 \\
       &       & 22.1 & $2.5$          &		 &         & 25 \\
000528 & \nodata & 11 & $2.3\pm 1.0$   & \nodata         & \nodata & 12 \\
000529 & \nodata & 9.0  & $2.8$        & \nodata         & \nodata & 26 \\
000926 & 2.037 & 54.9 & $2.23\pm 0.77$ & $-3.7\pm 1.5^a$   & 0.140   & 14,27 \\
       &       & 66.5 & $0.94\pm 0.14$ &                 &         & 27 \\
001025 & \nodata & 50.4 & $0.53\pm 0.10$ & $-3\pm 1.9^a$   & \nodata & 28 \\
001109 & \nodata & 19.3 & $7.1\pm 0.5$ & \nodata         & \nodata & 29 \\
010214 & \nodata & 7.7  & $6$          & $<-1.6$         & \nodata & 30 \\
       &       & 24.1 & $<0.5$         &                 &         & 30 \\
010220 & \nodata & 20.8 & 0.33         & $-1.2\pm 1.0$   & \nodata & 28 \\
010222 & 1.477 & 8.9  & $101\pm 11$    & $-1.33\pm 0.04$ & 0.080   & 14,31 \\
       &       & 32.7 & $18.7\pm 1.8$  &                 &         & 31 \\
       &       & 54.4 & $9.9\pm 0.5$   &                 &         & 31 \\
011211 & 2.14  & 11.0 & $1.9$          & $-1.7\pm 0.2$   & \nodata & 32 \\
020322 & \nodata & 18.8 & $3.5\pm 0.2$ & $-1.26\pm 0.23$ & \nodata & 33 \\
020405 & 0.698 & 41.0 & $13.6\pm 2.5$  & $-1.15\pm 0.95^d$ & 0.285 & 34,35,36 \\ 
020813 & 1.254 & 31.9 & $22$           & $-1.42\pm 0.05$ & 0.066   & 37,38 \\
021004 & 2.323 & 31.4 & $4.3\pm 0.7$   & $-1.0\pm 0.2$   & 0.240   & 39,40 \\
\enddata
\tablecomments{The columns are (left to right): (1) GRB name, (2)
redshift, (3) mid-point epoch of X-ray observation, (4) X-ray flux,
(5) temporal decay index ($F_X\propto t^{\alpha_X}$), (6) jet opening
angle, and (7) references for the X-ray flux and jet opening angle. 
$^a$ Due to the large uncertainty in the value of $\alpha_X$ we use
the median value for the sample, $\langle\alpha_X\rangle=-1.33\pm
0.38$.  $^b$ The redshift is based on matching the optical lightcurve
of SN\,1998bw to the red excess reported by \citet{bkd+99}.  $^c$ The
redshift limit is based on a detection of the afterglow in the optical
$U$-band \citep{jhb+01}.  $^d$ The inferred jet break is at $t=0.95$,
prior to the X-ray observation --- we use the model fit to extrapolate
the flux to $t=10$ hr (Berger et al. in prep.)} 
\tablerefs{(1) \citet{fag+98}; (2) \citet{pir01}; (3) \citet{fcp+98};
(4) \citet{paa+98}; (5) \citet{fks+01}; (6) \citet{mui+97}; (7)
\citet{slb+02}; (8) \citet{afa+99}; (9) \citet{mt98}; (10)
\citet{zaa+98}; (11) \citet{yfh+02}; (12) \citet{dpp+02}; (13)
\citet{naa+99}; (14) \citet{pk02}; (15) \citet{vgo+99}; (16)
\citet{faa+00}; (17) \citet{psa+01}; (18) \citet{nad+99}; (19)
\citet{fas+01}; (20) \citet{afv+00}; (21) \citet{fcd+99}; (22)
\citet{tmm+99}; (23) \citet{mtk+00}; (24) \citet{pfg+02}; (25)
\citet{apt+00}; (26) \citet{fgc+00}; (27) \citet{hys+01}; (28)
\citet{wro+02b}; (29) \citet{afp+00}; (30) \citet{fga+01}; (31)
\citet{zka+01}; (32) \citet{rwo+02}; (33) \citet{wro+02b}; (34)
\citet{pkb+02}; (35) \citet{mph02}; (36) Berger et al.~(in prep); (37)
\citet{pbg+02}; (38) \citet{vmf+02}; (39) Fox et al.~(in prep); (40)
Frail et al.~(in prep).} 
\end{deluxetable}

\clearpage

\begin{deluxetable}{cccccc}
\tablecolumns{6}
\tablewidth{0pc}
\tablecaption{X-ray Afterglow Data at $t=10$ hr\label{tab:xray2}} 
\tablehead {
\colhead {GRB}        		&
\colhead {$z$} 			&
\colhead {$F_{X,10}$}		&
\colhead {$L_{X,iso}$}		&
\colhead {$\theta_{\rm jet}$}	&
\colhead {$L_X$}		\\
\colhead {}			&
\colhead {}			&
\colhead {($10^{-13}$ erg cm$^{-2}$ s$^{-1}$)}	&
\colhead {($10^{45}$ erg s$^{-1}$)}	&
\colhead {}			&
\colhead {($10^{44}$ erg s$^{-1}$)}		
}
\startdata
970111 & \nodata & $3.36\pm 1.64$  & $2.56\pm 1.25$   & \nodata  &
\nodata \\
970228 & 0.695   & $27.50\pm 3.17$ & $6.82\pm 0.79$   & \nodata  &
\nodata \\  
970402 & \nodata & $2.86\pm 0.61$  & $2.18\pm 0.46$   & \nodata  &
\nodata \\
970508 & 0.835   & $9.60\pm 1.47$  & $3.74\pm 0.57$   & $0.391$  &
$2.82\pm 0.43$ \\
970815 & \nodata & $<18.47$	   & $<14.1$	      & \nodata  &
\nodata \\
970828 & 0.958   & $32.12\pm 6.31$ & $17.6\pm 3.4$    & $0.128$  &
$1.44\pm 0.28$ \\
971214 & 3.418   & $7.29\pm 0.87$  & $89.6\pm 10.8$   & $>0.100$ &
$>4.48$ \\
971227 & \nodata & $4.38\pm 1.26$  & $3.34\pm 0.96$   & \nodata  &
\nodata \\ 
980326 & $\sim 1$ & $<12.89$	   & $<9.82$	      & $<0.110$ &
$<0.59$ \\
980329 & \nodata & $10.68\pm 2.10$ & $8.14\pm 1.60$   & $0.081$  &
$0.27\pm 0.05$ \\
980515 & \nodata & $2.27\pm 0.90$  & $1.73\pm 0.69$   & \nodata  &
\nodata \\
980519 & \nodata & $6.14\pm 1.89$  & $4.68\pm 1.44$   & $0.040$  &
$0.04\pm 0.01$ \\
980613 & 1.096   & $7.03\pm 2.28$  & $5.36\pm 1.74$   & $>0.226$ &
$>1.36$ \\
980703 & 0.966   & $18.24\pm 4.97$ & $10.2\pm 2.8$  & $0.200$    &
$2.03\pm 0.55$ \\
981226 & \nodata & $6.19\pm 1.20$  & $4.72\pm 0.92$   & \nodata  &
\nodata \\
990123 & 1.600   & $66.09\pm 6.33$ & $128.31\pm 12.29$ & $0.089$ &
$5.08\pm 0.49$ \\
990217 & \nodata & $<1.25$	   & $<0.95$	       & \nodata &
\nodata \\
990510 & 1.619   & $41.07\pm 3.68$ & $82.09\pm 7.35$   & $0.054$ &
$1.20\pm 0.11$ \\
990627 & \nodata & $4.41\pm 0.85$  & $3.36\pm 0.65$    & \nodata &
\nodata \\
990704 & \nodata & $10.23\pm 3.34$ & $7.80\pm 2.54$    & \nodata &
\nodata \\
990705 & 0.840   & $3.11\pm 1.14$  & $1.23\pm 0.45$    & $0.096$ &
$0.06\pm 0.02$ \\
990806 & \nodata & $8.46\pm 3.14$  & $6.45\pm 2.39$    & \nodata &
\nodata \\
990907 & \nodata & $11.58\pm 6.95$ & $8.82\pm 5.29$    & \nodata &
\nodata \\
991014 & \nodata & $4.54\pm 1.71$  & $3.46\pm 1.30$    & \nodata &
\nodata \\
991216 & 1.020   & $287.21\pm 14.73$ & $183.22\pm 9.39$ & $0.051$ &
$2.38\pm 0.12$ \\
000115 & \nodata & $78.3\pm 14.12$ & $59.67\pm 10.76$  & \nodata &
\nodata \\
000210 & 0.846   & $4.56\pm 1.16$  & $1.83\pm 0.47$    & \nodata &
\nodata \\
000214 & \nodata & $10.25\pm 2.16$ & $7.81\pm 1.65$    & \nodata &
\nodata \\ 
000528 & \nodata & $2.61\pm 1.27$  & $1.99\pm 0.97$    & \nodata &
\nodata \\
000529 & \nodata & $2.43\pm 0.47$  & $1.85\pm 0.36$    & \nodata &
\nodata \\
000926 & 2.037   & $20.41\pm 8.06$ & $71.69\pm 28.31$  & $0.140$ &
$7.01\pm 2.77$ \\
001025 & \nodata & $67.85\pm 51.48$ & $51.71\pm 39.22$ & \nodata &
\nodata \\
001109 & \nodata & $17.02\pm 2.06$ & $12.97\pm 1.57$   & \nodata &
\nodata \\
010214 & \nodata & $3.95\pm 0.80$  & $3.01\pm 0.61$    & \nodata &
\nodata \\
010220 & \nodata & $0.79\pm 0.21$  & $0.61\pm 0.16$    & \nodata &
\nodata \\
010222 & 1.477   & $86.50\pm 9.88$ & $137.86\pm 15.75$ & $0.080$ &
$4.41\pm 0.50$ \\
011211 & 2.14    & $2.23\pm 0.39$  & $8.86\pm 1.56$    & \nodata &
\nodata \\
020322 & \nodata & $7.75\pm 0.67$  & $5.91\pm 0.51$    & \nodata &
\nodata \\
020405 & 0.698   & $68.98\pm 20.21$ & $17.29\pm 5.07$  & $0.285$ &
$6.98\pm 2.04$ \\
020813 & 1.254   & $113.98\pm 17.01$ & $121.21\pm 18.09$ & $0.066$ &
$2.61\pm 0.39$ \\
021004 & 2.323   & $13.50\pm 2.47$ & $65.36\pm 11.95$  & $0.240$ &
$18.7\pm 3.4$ \\
\enddata
\tablecomments{The columns are (left to right): (1) GRB name, (2)
redshift, (3) X-ray flux at $t=10$ hr, (4) X-ray luminosity at $t=10$
hr, (5) jet opening angle, and (6) beaming-corrected X-ray luminosity
at $t=10$ hr.} 
\end{deluxetable}

\clearpage

\clearpage

\end{document}